\documentclass[%
reprint,
amsmath,amssymb,
aps,
]{revtex4-1}

\usepackage{graphicx}
\usepackage{dcolumn}
\usepackage{bm}

\newcommand{\ket}[1]{|#1\rangle}

\begin{document}

\title{Physical qubit calibration on a directed acyclic graph}

\author{Julian Kelly}
\affiliation{Google}
\email{juliankelly@google.com}
\author{Peter O$'$Malley}
\affiliation{Google}
\author{Matthew Neeley}
\affiliation{Google}
\author{Hartmut Neven}
\affiliation{Google}
\author{John M. Martinis}
\affiliation{Google}

\date{\today}

\begin{abstract}
High-fidelity control of qubits requires precisely tuned control parameters. Typically, these parameters are found through a series of bootstrapped calibration experiments which successively acquire more accurate information about a physical qubit. However, optimal parameters are typically different between devices and can also drift in time, which begets the need for an efficient calibration strategy. Here, we introduce a framework to understand the relationship between calibrations as a directed graph. With this approach, calibration is reduced to a graph traversal problem that is automatable and extensible.
\end{abstract}

\maketitle

\tableofcontents

\section{\label{sec:intro}Introduction}

Computation on a quantum computer is realized by analog manipulations of large numbers of physical quantum bits (qubits). These control waveforms must be finely tuned as deviations from the ideal cause imperfect amplitudes and phases, inducing error. This presents a significant challenge as today's control hardware and qubit systems cannot be operated identically; each qubit's control must be calibrated individually, and ideal control parameters can drift in time. Furthermore, each qubit typically requires a host of different high fidelity operations, such as single-qubit rotations, multi-qubit gates, measurement, and reset for a digital quantum computer~\cite{divincenzo2000physical, fowler2012surface}. The field of quantum optimal control ~\cite{werschnik2007quantum, shapiro2003principles, khaneja2005optimal} lies at the center of this challenge, where precise control over quantum systems has been demonstrated in state transfer~\cite{bardeen1997feedback}, high-fidelity logic operations~\cite{sporl2007optimal, egger2014adaptive, kelly2014optimal, mcclure2016rapid, rol2017restless}, combating control parameter drift~\cite{koch2004stabilization, kelly2016scalable}, and macroscopic quantum systems~\cite{hohenester2007optimal}. However, quantum optimal control is typically applied to a single operation, such as calibrating a particular logic gate or quantum state. In this work, we focus on a broader challenge: how do we navigate the full suite of tasks required for a quantum computer, from initial device bring-up to logic operation calibration to algorithmic control?

Typically, control parameters for a device are determined by following a carefully chosen sequence of calibration experiments, where the result of one experiment generally feeds into the next. The simplest strategy is to sequence through these experiments from start to end, in order. However, this breaks down with the introduction of parameter drift, the need for debugging, or the desire to reconfigure the system on-the-fly. Re-starting the sequence from the beginning repeatedly is non-ideal given the considerable time expense of the calibration sequence. Simply put, there is a need to move fluidly both forwards and backwards through a sequence of experiments, and for a way to represent the relationship between calibration experiments.

For small systems, complex calibrations and parameter drift can be handled by careful, manual tuning. Here, an adept user can navigate forwards and backwards through experiments for a handful of qubits, as they have learned the relationships between experiments through experience. It is this experience that provides inspiration for our work.

Naturally, manual control is not scalable, so a fully autonomous solution is desireable to operate systems with more than a few dozen qubits. Our goal for this paper is to provide a high-level description of the interplay between calibration experiments, and how this can be used to identify and maintain control parameters for a quantum computer.

\section{\label{sec:calproblem}The Calibration Problem}

We define the calibration problem as follows:  \textit{What is the optimal way to identify and maintain control parameters for a system of physical qubits given incomplete system information?} To begin, we define the following terms:
\begin{description}

\item[Parameter] A qubit control parameter, such as the driving duration of a pi-pulse. 

\item[Experiment] A collection of static control waveforms and measurements where no parameters are varied. This may correspond to, for example, a gate sequence and measurement to determine the output probability distribution or a gate sequence and tomography to interrogate the state of the qubit(s). Typically each experiment is repeated a number of times to gather statistics. 

\item[Scan] A collection of experiments where one or more parameters are varied. For example, a rabi driving scan where the length of the drive pulse is changed for each individual experiment. Scans can also be more complex, e.g. multi-parameter optimizations. 

\item[Figure of merit] A number measured using a scan or experiment used to quantify how well a qubit is operating. 

\item[Tolerance] A threshold on a figure of merit for determining whether the control parameters are determined well enough. For example, we may specify that a pi pulse should be within $10^{-4}$ radians of a $\pi$ rotation.

\item[Spec] A figure of merit is either within tolerance (in spec) or not (out of spec). 

\end{description}

\subsection{\label{sec:cals}Calibrations (Cals)}
We define a \textbf{calibration (cal)} as a procedure used to determine a new value for one or more parameters. A cal consists of the following components:

\begin{enumerate}
\item A set of parameters that are targets of cal. 
\item A scan used to generate experimental data relevant to the parameters.
\item Functions used to analyze the experimental data to determine figures of merit. The figures of merit are used to determine if the parameters are in spec, and what the optimal parameter values are. 
\end{enumerate}

\subsection{\label{sec:loop}The Calibration loop}
Cals share a generic structure, as follows:
\begin{enumerate}
\item Run scan.
\item Analyze data to determine optimal parameter values.
\item Update parameters.
\end{enumerate}

In fully calibrating a system of qubits, this procedure is followed through a collection of different cal procedures across all of the qubits. 

\begin{figure}[!]
\centering
\includegraphics[width=0.48\textwidth]{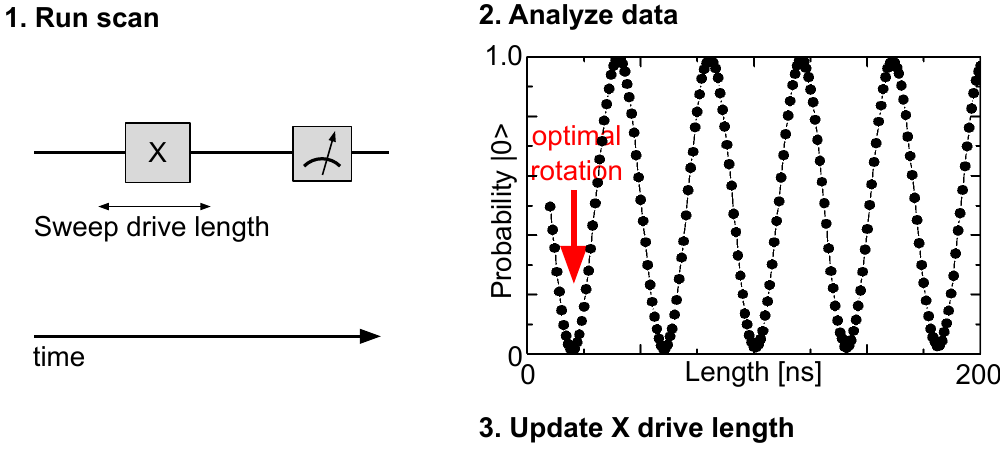}
\caption{\textbf{An example of a rabi driving cal.} (1) The rabi driving scan is performed, consisting of a collection of experiments, where each experiment has a single drive length, and the average probability of the $\ket{0}$ state is measured. (2) The data is analyzed, and the optimal drive length is determined. (3) The qubit parameter for the driving length of an X pulse is updated to the optimal value.}
\label{fig:rabi}
\end{figure}

\subsection{\label{sec:dep}Dependencies}

Much of the complexity in determining control parameters arises because calibration experiments can \textit{depend} on one another. The process of calibrating a qubit uses bootstrapping, where parameters discovered from simple cals are fed into more sophisticated cals which enhance the control capabilities of a qubit beyond what could be done initially. In the rabi driving example (Figure~\ref{fig:rabi}), this scan can only be performed once we know the frequency at which to drive the qubit. In turn the qubit frequency cal can only be performed once we know how to measure the state of the qubit. In general, a cal can depend on parameters derived from other cals, which can themselves depend on other cals. This is especially tricky in the presence of drift: a parameter that drifts from its calibrated value during subsequent cals can poison the data.  

\section{\label{sec:optimus}Optimus}

Our goal is to build a system to solve the calibration problem with the following properties: 
\begin{enumerate}
\item Calibrates a qubit system automatically from scratch (from its unknown initial state). 
\item Chooses the sequence of calibrations to perform. 
\item Takes a minimal amount of wall clock time. 
\item Handles parameter drift, i.e. has a strategy for detecting drift and revisiting calibrations that may have drifted.
\item Detects if a calibration is working improperly via self-diagnosis.
\end{enumerate}

\subsection{\label{sec:graph}The calibration graph}

\begin{figure}[!]
\centering
\includegraphics[width=0.48\textwidth]{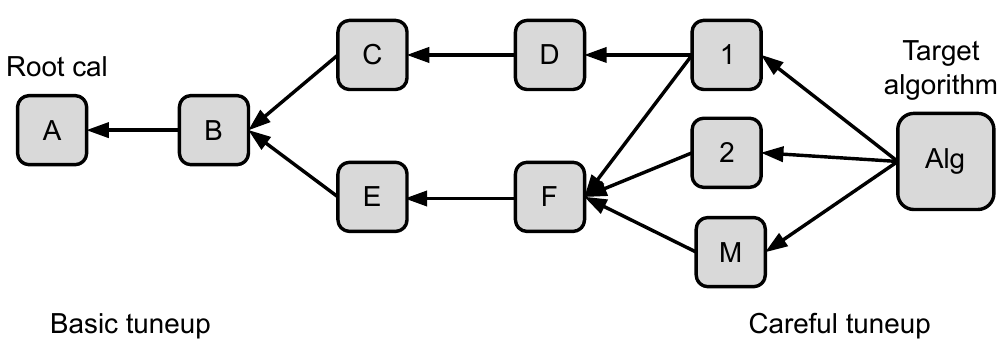}
\caption{\textbf{Calibrations in a directed acyclic graph.} As an example, we have an algorithm that we would like to run that depends on high fidelity single-qubit gates (1), two-qubit gates (2) and measurement (M). In turn, these cals have a graph of dependencies which extend back to the root.}
\label{fig:graph}
\end{figure}

We introduce our approach to the calibration problem, named Optimus, which satisfies these properties. The essential insight of Optimus is to formulate the dependency relationships between calibrations as a directed acyclic graph (DAG). Each cal is represented by a node in the graph, and each dependence between cals is a directed edge. The direction of the edge denotes which cal depends on the other. For example, in Figure~\ref{fig:graph} we see that cal B depends on cal A, while cal 1 depends on cals D and F. Cal A is a root cal; it does not depend on any other cals and so can be run on a qubit system in an otherwise unknown state.

\subsection{\label{sec:knowledge}Where does system knowledge live?}
So far, we have only outlined a generic control system, and have not touched on where specific system information should live. Any person who has spent extensive time working with qubits has built up significant system specific knowledge, such as roughly what calibrations should be run in what order to bring up a qubit, what data for various experiments should look like, and what typical operation parameters are. So where does this information live?

The Optimus framework allows one to codify all of that lab experience explicitly into the graph in a useful way. The dependency structure of the graph, which experiments are used on each node, the tolerances for those experiments, analysis functions, etc all represent our knowledge of how the system works. We expect that these specifics will be different from a superconducting qubit system to that of ion traps or quantum dots, or even a different design of superconducting qubit. Even among one particular system there is some freedom in how the graph is constructed, and this is a good thing. We may use different graphs for different purposes, such as standard characterization vs high fidelity algorithm operation.

By putting all of this system knowledge into the calibration graph, multiple users can contribute to the collective knowledge of the best operating procedure. In manual operation, it takes time for new users to become familiar with all of the elements of system calibration, and users of different experience levels typically handle the system differently. Now, the graph represents the communal best working knowledge of how to solve the calibration problem. If one user finds a better way to write a node or structure the graph, they can systematically improve the calibration routine and distribute that to other users as the nodes are modular. The graph is also useful as a pedagogical tool, as it explicitly describes the relationships between calibrations, analysis functions, and thresholds. 

\subsection{\label{sec:dag}The calibration problem on a DAG}

\begin{figure}[!]
\centering
\includegraphics[width=0.48\textwidth]{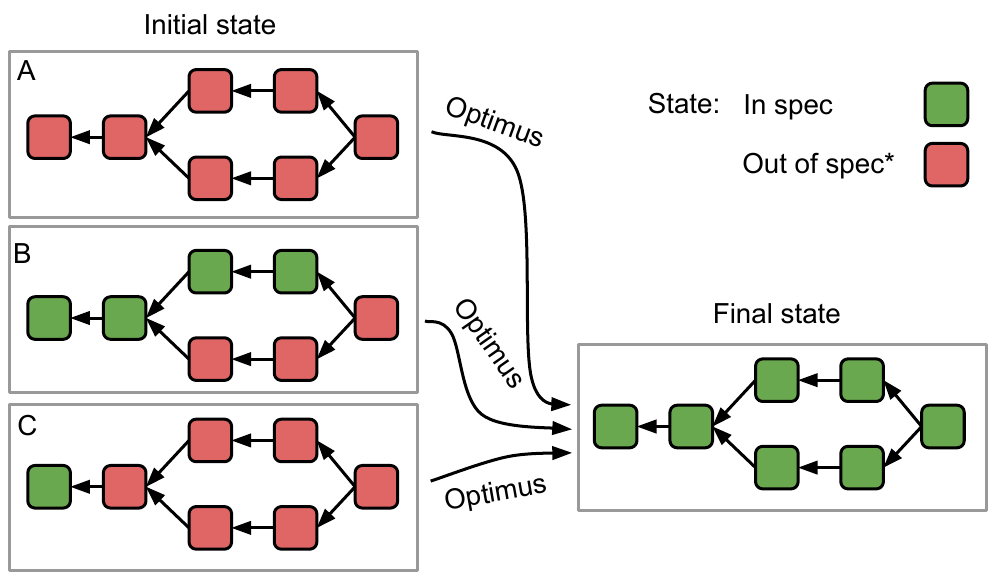}
\caption{\textbf{The calibration problem can be formulated in terms of a DAG.} The dependency arrows point towards the root. The goal of the calibration problem is to take systems with little information (A) or partial information (B, C) and bring them to the state where all cals are in spec. *Note that cals where the state is unknown (e.g. never previously measured) are assumed to be out of spec.}
\label{fig:anystate}
\end{figure}

In addition to the graph structure, we also introduce the state, which represents all current knowledge of the system based on previously acquired data. This knowledge is similar to an experimentalists lab notes: it contains information on which cals are working and when they were last run, for example. Each cal can be in a different state: ``in spec'' denoting that the figures of merit associated with the cal are in spec, or ``out of spec'' if they are out of spec. For example if we just successfully ran a cal we would mark it as in spec; if it failed we would mark it as out of spec; finally, if the state is unknown we would assume it is out of spec until validated. By definition, a cal cannot be in spec if it has a dependency out of spec; we address how to determine the state of the system in section~\ref{sec:state}. With this formalism, we can restate the calibration problem in terms of the DAG, as seen in Figure~\ref{fig:anystate}. 

To help combat parameter drift, we also associate a timeout period with each cal, and record the time when the cal was last run. The timeout period is the characteristic drift time, or the timescale that we expect that the parameters associated with the cal should be validated or updated. Note that these timeout periods can be experimentally determined by processing data generated from system operation.  

\subsection{\label{sec:methods}Interacting with calibrations}

There are a number of ways of interacting with calibrations, depending on what information we are trying to learn. In an effort to reduce the amount of time Optimus consumes (the cost of calibrations), we introduce three methods, \texttt{check\_state}, \texttt{check\_data}, and \texttt{calibrate}. These three functions gate expensive calibrations that may not be needed: \texttt{check\_state}, \texttt{check\_data}, and \texttt{calibrate} acquire no data, little data, and lots of data respectively. Importantly, only \texttt{calibrate} (which acquires significant amounts of data) is used to update parameter values, as it provides the highest confidence. 

\subsubsection{\label{sec:state}check\_state}
\texttt{check\_state} answers the question, ``Based on our prior knowledge of the system, and without experiments, are we confident the figures of merit associated with this cal are in spec?'' The purpose of \texttt{check\_state} is to help higher level algorithms determine where to allocate resources running experiments.

\texttt{check\_state} should report a pass if and only if the following are satisfied:
\begin{enumerate}
\item The cal has had \texttt{check\_data} or \texttt{calibrate} pass within the timeout period.
\item The cal has not failed \texttt{calibrate} without resolution. 
\item No dependencies have been recalibrated since the last time \texttt{check\_data} or \texttt{calibrate} was run on this cal. 
\item All dependencies pass \texttt{check\_state}. 
\end{enumerate}

\subsubsection{\label{sec:checkdata}check\_data}
The purpose of \texttt{check\_data} is to experimentally determine the state of the node, while running a minimal number of experiments. \texttt{check\_data} answers two questions, ``Is the parameter associated with this cal in spec, and is the cal scan working as expected?''

\begin{figure}[!]
\centering
\includegraphics[width=0.48\textwidth]{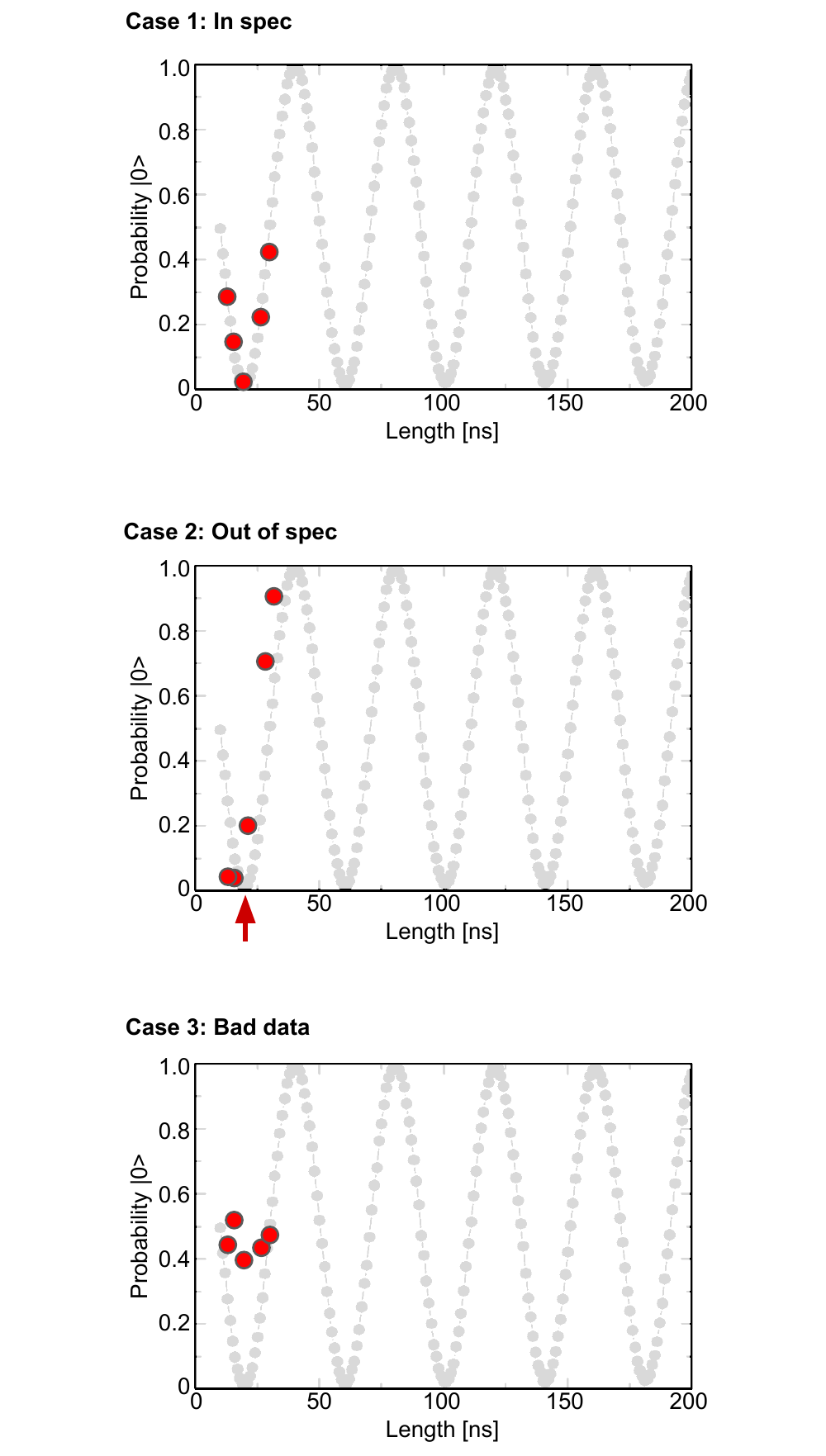}
\caption{\textbf{\texttt{check\_data} uses a minimal number of points to distinguish between three cases.} Case 1: the data lies on top of the expected data indicating the parameter is in spec. Case 2: the data are offset from the expected data, indicating the parameter is not in spec. Case 3: The data is noise, indicating a dependency is bad.}
\label{fig:checkdata}
\end{figure}

This can be understood with a Rabi driving cal example, as shown in Figure~\ref{fig:checkdata}. \texttt{check\_data} takes a minimal amount of data (the red points), which we superimpose onto the black dots, representing what we expect the data to look like. In the first case, we see that the red dots lie within acceptable tolerance of the black dots indicating that our parameter matches the expectation for the experiment and is in spec. In the second case, we see that there is a shift between where the expected and actual minimal of the curve are, indicating the parameter is out of spec (which is recorded in the state). In the third case, we see that the data looks like noise instead of lying on a curve, indicating that the scan for this cal is not working and returning bad data. When we receive bad data, it likely indicates that a dependence for this cal has a bad state.

\subsubsection{\label{sec:calibrate}calibrate}
\texttt{calibrate} is the canonical calibration loop described above. We allow \texttt{calibrate} to take as much  data as needed to determine the optimal value for a parameter. We then analyze the data and update the parameter associated with the cal. Additionally, we verify that the data generated by the calibration scan are not bad data. In the case of bad data, we generate an error as bad data should have been caught previously by \texttt{check\_data}. 

\subsection{\label{sec:attributes}Calibration attributes}

For completeness, we elaborate on what constitutes a calibration node:
\begin{enumerate}
\item Parameter(s) that are the target of cal. 

\item Scans used to generate experimental data relevant to the parameter(s).
\begin{enumerate}
\item \texttt{check\_data} scan (minimal data).
\item calibration scan (more data).
\end{enumerate}

\item Helper functions for analysis.
\begin{enumerate}
\item \texttt{check\_data} analysis.
\item \texttt{calibration} analysis.
\item Supplementary checks, e.g. for qubit parameter inconsistencies.
\end{enumerate}

\item Tolerances for a figure of merit. 
\item Timeout period.
\end{enumerate}

\subsection{\label{sec:traversal}Graph traversal}
With the specifics of how cals are defined and how we can interact with them, system calibration is now reduced to a graph traversal problem. So, we need algorithms to determine which cals to run, and the order they will execute. 

With this in mind, we introduce two algorithms for navigating the DAG: \texttt{maintain} and \texttt{diagnose}. The specific nature of these algorithms is to minimize the amount of time is the required to bring the system back to a good state. 

\subsubsection{\label{sec:maintain}Maintain concept}
\texttt{maintain} is the primary interface for Optimus, and is the highest level algorithm. We call \texttt{maintain} on the cal that we want in spec, and \texttt{maintain} will call all necessary subroutines to get that node in spec. The goal of \texttt{maintain} is to only begin acquiring data on the node closest to root that fails \texttt{check\_state}. We do this as we want to begin changing parameters on the least dependent calibration and work forwards from there. 

\begin{figure}[!]
\centering
\includegraphics[width=0.48\textwidth]{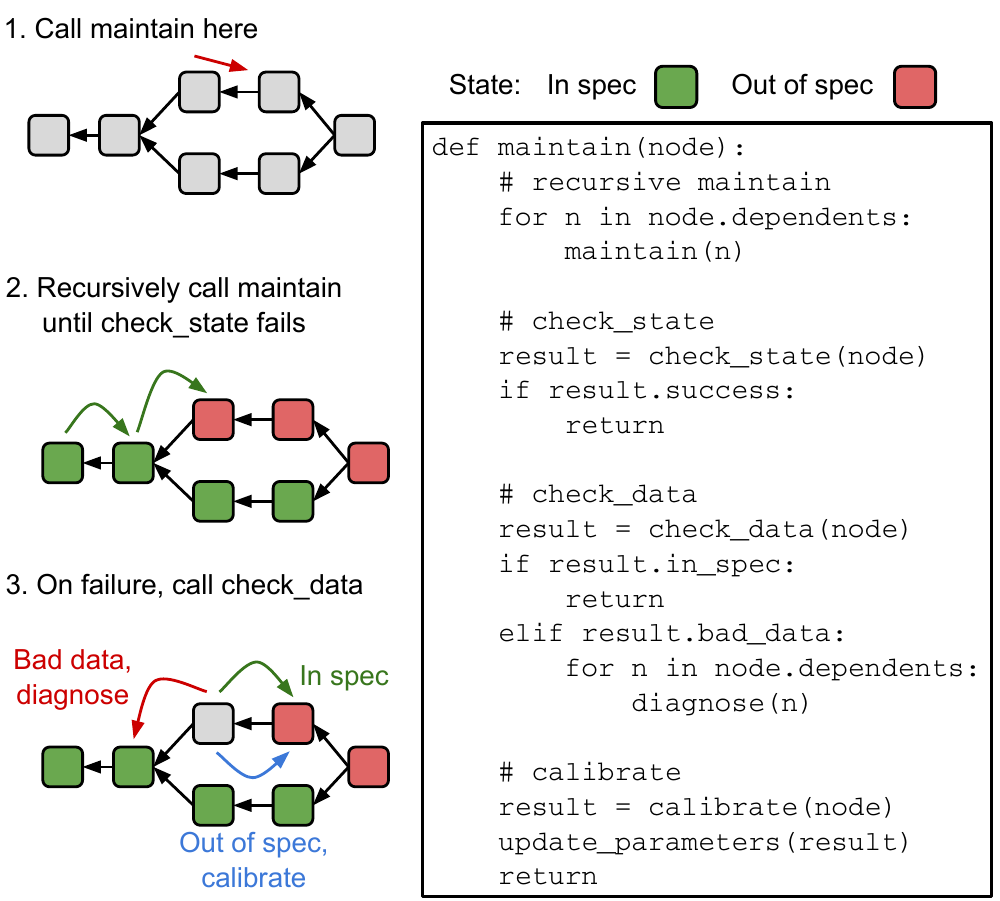}
\caption{\texttt{maintain} is called on the cal we want in a good state, and is called recursively down to the root node. The node closest to the root which fails \texttt{check\_state} runs \texttt{check\_data}, and decide to proceed, \texttt{calibrate}, or call \texttt{diagnose} depending on the outcome. We iterate recursively through the graph with this strategy.}
\label{fig:maintain}
\end{figure}

At the first node that fails \texttt{check\_state}, the \texttt{maintain} algorithm will begin interrogating state of the node experimentally with \texttt{check\_data}. Depending on the outcome, it will either proceed (in spec), \texttt{calibrate} and proceed (out of spec), or call \texttt{diagnose} (bad data), see Fig~\ref{fig:maintain}. It will then continue iterating through the graph using this basic strategy.

\subsubsection{\label{sec:diagnose}Diagnose concept}

\begin{figure}[!]
\centering
\includegraphics[width=0.48\textwidth]{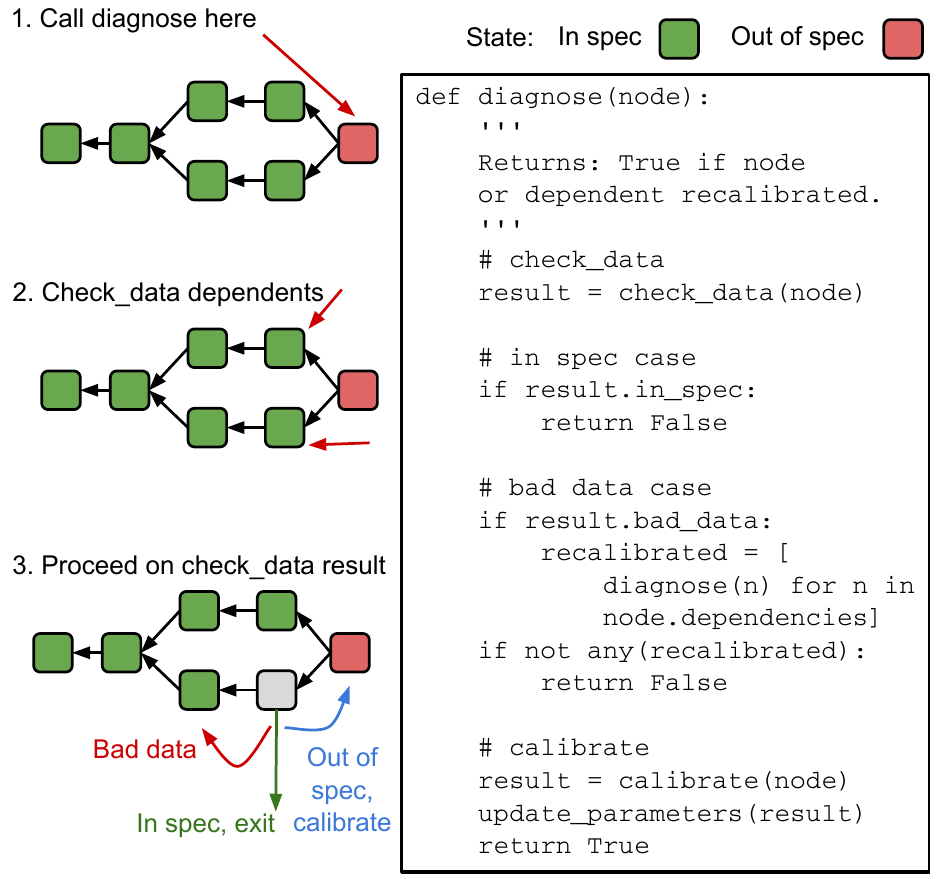}
\caption{Diagnose is called on a cal that has returned bad data from \texttt{check\_data}. We then investigate each of the dependents. If a dependent is in spec, we continue to other dependents of the original cal. If it is out of spec, we \texttt{calibrate} and proceed towards the original cal. If we get bad data, we recursively \texttt{diagnose}.}
\label{fig:diagnose}
\end{figure}

\texttt{diagnose} is a separate algorithm that is called by \texttt{maintain}, in the special case that \texttt{check\_data} identifies bad data. The \texttt{diagnose} algorithm represents a strategy shift from \texttt{maintain}; \texttt{diagnose} makes no calls to \texttt{check\_state} and only makes decisions based on data returned from \texttt{check\_data} and \texttt{calibrate}, see Fig~\ref{fig:diagnose}. This can be understood by realizing that \texttt{maintain} assumes that our knowledge of the state of the system matches the actual state of the system. If we knew a cal would return bad data, we wouldn't dedicate experiments to it. \texttt{diagnose} is only invoked when we have experimentally determined there is a mismatch between the actual system state and our knowledge of the system state. So, the purpose of \texttt{diagnose} is to repair inaccuracies in our knowledge of the state of the system, so that \texttt{maintain} can resume. 

\subsection{\label{sec:time}Saving time}

We design \texttt{maintain} to start in the optimal location (to the best of our knowledge) to avoid extra work. For example, if a node times out, so that we can no longer consider it to be in spec, then all higher-level nodes that depend on it (directly or indirectly) are also out of spec. So, to bring the system back into a good state, we want to begin with the lowest-level node that we determine to be out of spec and work upwards from there. This minimizes the work we have to do because we should not bother trying to tune up a higher level node when we know a lower-level dependency is not in spec.

Similarly, switching directions in the \texttt{diagnose} phase after we find bad data is also a way to avoid doing extra work. When we find that the data for a particular node is bad, even though we believe that its dependencies are in spec, we first check the immediate dependencies more carefully. If these can be brought back into spec, then we are all good and can continue. We don't immediately jump to the conclusion that every dependent node all the way back to the root is bad, which would cause us to start over and do a lot of extra work. So we switch directions and work from the top down in the \texttt{diagnose} phase.

\subsection{\label{sec:acyclicity}Note on acyclicity}

\begin{figure}[!]
\centering
\includegraphics[width=0.48\textwidth]{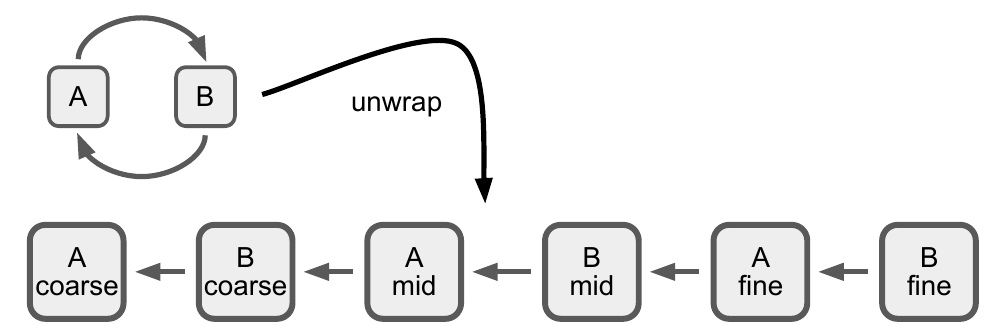}
\caption{Acyclicity can be removed by unwrapped cyclic dependencies into layers of precision. This allows us to iterate back and forth updating parameters until they are both finely tuned.}
\label{fig:acyclicity}
\end{figure}

Acyclicity is a desirable property for the graph to have, as it makes the graph traversal algorithms easier to build and control. However, our systems are not naturally acyclic. To handle this, we can unwrap a cyclic dependence into layers of precision. For example, as in Figure~\ref{fig:acyclicity} if A and B both depend on each other, we can unwrap it into layers of coarse, mid, and fine calibration. By doing this, we essentially iterate through parameter space until we achieve the desired level of precision. Alternatively, in some cases we can design a cal scan that simultaneously optimizes both parameters. Deciding between these cases is an element of the overall DAG design.

\subsection{\label{sec:errors}Handling Errors}
In some cases, we may raise errors when encountering unexpected behavior. For example, we raise a DiagnoseError if we find that \texttt{diagnose} was invoked, but no dependencies were out of spec. In this case (assuming we have defined tolerances that are physically achievable), it is possible the device is behaving in a non-ideal manner, or the DAG does not accurately represent the system behavior. We can try bringing up the device in a different operating condition, or mark it as bad. 

\subsection{\label{sec:mq}Multi-qubit calibrations}
System tune-up requires multi-qubit calibrations in addition to single-qubit calibrations. Ensuring these behave as expected involves unifying methods such as \texttt{check\_state} to to work in the following way: if one qubit fails \texttt{check\_state} due to some dependency but the other succeeds, the multi-qubit node should fail \texttt{check\_state}. These issues can be addressed in the specific software implementation.

\subsection{\label{sec:extensibility}Extensibility to other physical systems}
We have written Optimus with control over physical qubit systems for quantum computing. However this approach is generically applicable to a variety of platforms that require careful calibration of a physical system.

\section{\label{sec:outlook}Outlook}
We have presented Optimus as a solution to the calibration problem, where we reformulated the suite of calibration scans as a directed acyclic graph. We then defined a number of methods of interacting with different cals as a way to minimize the number of experiments require to fully calibrate a system. Doing this allows us to use graph traversal strategies to tackle the calibration problem. We have been using Optimus to successfully automate the calibration of multi-qubit systems.

\addcontentsline{toc}{chapter}{Bibliography}
\bibliographystyle{apsrev}
\bibliography{bibliography}

\end{document}